\documentclass[prl,10pt,twocolumn,superscriptaddress,floatfix,nofootinbib,showpacs]{revtex4-1}

\usepackage[utf8]{inputenc}  
\usepackage[T1]{fontenc}     
\usepackage[british]{babel}  
\usepackage[sc,osf]{mathpazo}
\usepackage[scaled=0.86]{berasans} 
\usepackage[colorlinks=true, citecolor=blue, urlcolor=blue]{hyperref}  
\usepackage{scrextend}
\usepackage{graphicx} 
\usepackage[babel]{microtype}  
\usepackage{amsmath,amssymb,amsthm,bm,amsfonts,mathrsfs,bbm} 

\usepackage{xspace}  
\usepackage{pgf,tikz}
\usepackage{xcolor,colortbl}
\usepackage{array}
\usepackage{bigstrut}
\usepackage{braket}
\usepackage{color}
\usepackage{natbib}

\newcommand{\be}{\begin{equation}}
\newcommand{\ee}{\end{equation}}
\newcommand{\ba}{\begin{eqnarray}}
\newcommand{\ea}{\end{eqnarray}}

\begin{document}


\title{Quantum nonlocality does not demand all-out randomness in measurement choice}

\author{Manik Banik}
\affiliation{Optics \& Quantum Information Group, The Institute of Mathematical Sciences, HBNI, C.I.T Campus, Tharamani, Chennai 600 113, India.}

\author{Samir Kunkri}
\affiliation{Mahadevananda Mahavidyalaya, Monirampore, Barrackpore, North 24 Parganas-700120, West Bengal, India.}

\author{Avijit Misra}
\affiliation{Optics \& Quantum Information Group, The Institute of Mathematical Sciences, HBNI, C.I.T Campus, Tharamani, Chennai 600 113, India.}

\author{Some Sankar Bhattacharya}
\affiliation{Physics and Applied Mathematics Unit, Indian Statistical Institute, 203 B. T. Road, Kolkata 700108, India.}

\author{Arup Roy}
\affiliation{Physics and Applied Mathematics Unit, Indian Statistical Institute, 203 B. T. Road, Kolkata 700108, India.}

\author{Amit Mukherjee}
\affiliation{Physics and Applied Mathematics Unit, Indian Statistical Institute, 203 B. T. Road, Kolkata 700108, India.}

\author{Sibasish Ghosh}
\affiliation{Optics \& Quantum Information Group, The Institute of Mathematical Sciences, HBNI, C.I.T Campus, Tharamani, Chennai 600 113, India.}

\author{Guruprasad Kar}
\affiliation{Physics and Applied Mathematics Unit, Indian Statistical Institute, 203 B. T. Road, Kolkata 700108, India.}


\begin{abstract}
Nonlocality is the most characteristic feature of quantum mechanics. John Bell, in his seminal 1964 work, proved that \emph{local-realism} imposes a bound on the correlations among the measurement statistics of distant observers. Surpassing this bound rules out local-realistic description of microscopic phenomena, establishing the presence of nonlocal correlation. To manifest nonlocality, it requires, in the simplest scenario, two measurements performed randomly by each of two distant observers.
In this work, we propose a novel framework where three measurements, two on Alice's side and one on Bob's side, suffice to reveal quantum nonlocality and hence does not require all-out randomness in measurement choice. Our method relies on a very naive operational task in quantum information theory, namely, the minimal error state discrimination. As a practical implication this method constitutes an economical entanglement detection scheme, which uses a less number of entangled states compared to all such existing schemes. Moreover, the method applies to class of generalized probability theories containing quantum theory as a special example. 
   
\end{abstract}


\maketitle

\emph{Introduction.} Entanglement is one of the most non-classical manifestations of multipartite quantum system \cite{Horodecki'09}. It has revealed  puzzling features of quantum theory (QT) since its early advents. Whereas Einstein, Podolsky, and Rosen used entanglement to show incompleteness of QT \cite{Einstein'35}, Bell used the same to rule out the possibility of any \emph{local-realistic} description underlying QT \cite{Bell'64}. This nonlocal behavior of correlations arising from quantum states, undoubtedly, shows one of the most fundamental departures of the theory from its classical counterpart. However, this nonlocal correlation has no contradiction with relativistic causality principle. 
Bell's result provides an empirical method to demonstrate the presence of such correlations, which has already been successfully tested in number of experiments \cite{Aspect'82,Groblacher'07,Hensen'26}. In the simplest such test, proposed by Clauser, Horne, Shimony and Holt (CHSH), one has to consider two spatially separated observers, each performing two different measurements at random \cite{CHSH'69}. Under the assumption that the state vector provides a complete description of physical system, then following a reasoning of Einstein presented in 1927 Solvay conference, one can construct a simple argument to establish nonlocal behavior of QT, without invoking Bell's theorem \cite{Bacciagaluppi'09,Norsen'05}. In the recent past such kind of argument has been extended for \emph{$\psi$-ontic} ontological models \cite{Harrigan'10} and for non maximally \emph{$\psi$-epistemic} ontological models \cite{Leifer'13,Banik'14} of QT. The novelty of Bell's theorem over all such arguments is that it is not limited to a particular class of ontological models. Furthermore, it provides an empirical way to test the nonlocal behavior and also gives operational mean to quantify nonlocality.           
\par
In this work we propose a new method to establish nonlocal behavior of bipartite quantum states. Our method is based on an operational task: a guessing game between two spatially separated players, Alice and Bob namely,
who share a bipartite quantum state. Alice is supposed to prepare two different decompositions of Bob's marginal state by performing suitable local measurements on her part. Bob performs a single measurement and uses the measurement statistics to calculate a payoff function. This payoff function turns out to be the success probability of binary minimal error state discrimination task. We show that whenever the payoff value overcomes a threshold it detects nonlocality of the shared bipartite state. It is important to note that all-out randomness in measurement choice is not necessary to exhibit nonlocality as Bob performs only one measurement. This makes our method experimentally less demanding compared to the simplest Bell scenario. 
\par 
On practical ground our work provides a novel scheme of verifying quantum entanglement which with advent of quantum information theory has been identified as useful resource for many information processes- starting from the canonical ones like quantum cryptography \cite{Ekert'91}, quantum teleportation \cite{Bennett'93}, quantum dense coding \cite{Bennett'92} to very recent satellite-based quantum communication network \cite{Yin'17}. Compared to all the existing entanglement verification schemes, our method is advantageous as it requires less entangled pairs at the verification stage. It also embodies a generic theoretical appeal as we find that the protocol can be studied in the framework of generalized probabilistic theories (GPT) which contains classical probability theory and QT as special cases.     

\emph{Rudiment of GPT.} This framework capture all probabilistic theories that contains concept of states and effects to make predictions for probabilities of measurement outcomes. Though the original formulation dates back to sixties \cite{Mielnik'68,Ludwig'67,Mackey'63}, it gets renewed interest in the recent past aiming to derive QT from physical/information theoretic principles \cite{Hardy'01,Barrett'07,Masanes'11,Chiribella'11}. Classical probability theory and QT are contained within this framework as special cases. However, the framework can allow more general theories containing post-quantum correlations \cite{Popescu'94}. 
\par
A physical system $S$ is described by some state $\omega$ belonging to some state space $\Omega_S$ which is a compact and convex set embedded in some real vector space $V$.
Convexity of $\Omega_S$ assures any statistical mixture of states as a valid state. The extremal points of the set $\Omega_S$ that do not allow any decomposition in terms of other states are called pure states or states of maximal knowledge.           
A general mapping from states to probabilities is described by effects, $e:\Omega_S\mapsto[0,1]$. Collection of all effects, denoted as $\mathcal{E}_S$, forms a convex set embedded in the vector space $V^*$, dual to the vector space $V$. We denote the probability of occurrence of some effect $e_x$ on some state $\omega_y$ as $p(x|y):=e_x(w_y)$. Unit effect $u$ is introduced with the property that, $u(\omega)=1,~\forall~\omega\in\Omega_S$. An $n$-outcome measurement $M_n\in\mathcal{M}_n$ is a collection of $n$ effects $M_n:=\{e_x|\sum_xe_x=u\}_{x=1}^n$.
This framework also considers composite systems with local state spaces (say) $\Omega_1$ and $\Omega_2$. Such a composition must be constructed in accordance with no-signaling (NS) principle that prohibits instantaneous communication between two spatially separated locations. Under another less intuitive assumption called tomographic locality \cite{Hardy'01}, the state space of the composite system  lives in the vector space $V_1\otimes V_2$. We denote the composite state space as $\Omega=\Omega_1\otimes\Omega_2=(V_1\otimes V_2)_+^1$, where $(V_1\otimes V_2)_+^1$ denotes the normalized positive cone with normalization is given by the order unit $u_1\otimes u_2\in V^*_1\otimes V^*_2$. There is no unique choice for the positive cone, but it lies within the two extremals, $(V_1\otimes_{min} V_2)_+:=\{\sum\alpha_{ij}\omega^i_1\otimes\omega^j_2|\alpha_{ij}\in\mathbb{R}_+, \omega^i_1\in (V_k)_+\}$ and $(V_1\otimes_{max} V_2)_+:=(V^*_1\otimes_{min} V^*_2)^*_+$. Only when local state spaces are simplexes, i.e., for the case of classical probability theory, the choice for tensor product is unique \cite{Namioka'69}. 
\par
The Hilbert space formulation of QT lies within this framework. State of a quantum system $S$ associated with a Hilbert space $\mathcal{H}_S$ is describe by density operator $\rho\in\mathcal{D}(\mathcal{H}_S)$, where $\mathcal{D}(\mathcal{H}_S)$ is the convex set of hermitian, positive, and trace one operators acting on $\mathcal{H}_S$. Measurement is describe by positive-operator-valued-measure, $M:=\{E_k~|~E_k\ge 0,~\sum_kE_k=\mathbf{1}\}$ and the outcome probability is given by the Born rule, i.e., $\mbox{tr}[\rho E_k]$. The composite system is described by quantum mechanical tensor product which is neither minimal nor maximal tensor product space, rather lies strictly in between.
\par
GPT framework provides only an operational description of the events performed in laboratory, but does not tell anything about the \emph{reality} of the physical system. The question regarding reality of physical system can be well addressed in the ontological model of an operational GPT \cite{Spekkens'05,Leifer'14}. An ontological model underlying a GPT is specified by a triplet $(\Lambda,\mu,\xi)$, with $\Lambda$ being the space of possible ontic states for the physical system. Every operational preparation $\omega$ gives a probability distribution $\mu_{\omega}(\lambda)$ over the ontic state space $\Lambda$, i.e.,  $\mu_{\omega}(\lambda)\ge 0,~\forall~\lambda,\omega$ and $\int_{\lambda\in\Lambda}d\lambda\mu_{\omega}(\lambda)=1,~\forall~\omega$. Ontic states specify outcome probability of $e_k$ in a measurement $M=\{e_k\}_{k=1}^n$ by some response function $\xi_{e_k|M}(\lambda)\in[0,1]$ which satisfies $\sum_k\xi_{e_k|M}(\lambda)=1$. The ontological model
must reproduce the statistical predictions of the GPT, i.e., $\forall~e_k, \omega:~\int_{\lambda\in\Lambda}d\lambda\mu_{\omega}(\lambda)\xi_{e_k|M}(\lambda)=e_k(\omega)$. Certain combination of assumptions on the ontological model may not be compatible with the operational statistics of GPT. The no-go results by Bell \cite{Bell'64} and Kochen-Specker \cite{Kochen'67} are seminal such examples that exclude classes of ontological model for operational QT. In what follows, we discuss
a familiar quantum protocol, namely minimal error state discrimination \cite{Helstrom'69} and a related no-go result derived under certain assumption on ontological level \cite{Schmid'17}.

\emph{Schmid-Spekkens no-go result.} A characteristic feature of quantum mechanics is that no measurement can perfectly discriminate between two given non-orthogonal states. However, 
these states can be discriminated imperfectly in the following ways: unambiguous state discrimination and  minimum-error state discrimination.  
In this work, we focus on the latter,
which has also been studied in the GPT framework of late \cite{Kimura'09,Nuida'10,Bae'11}. 
\par
Suppose, Alice is randomly given one of the states $\{\omega_x\}_{x=1}^k\subset\Omega$, chosen with a prior probability distribution $\{p_x~|~p_x\ge 0,~\sum_xp_x=1\}_{x=1}^k$. She has to guess the state optimally. Alice performs a measurement $M_n=\{e_x~|~\sum_xe_x=u\}\in\mathcal{M}_n$ to optimize her success probability of guessing, $P^{\Omega}_S=\sup_{M_n\in\mathcal{M}_n}\sum_{x=1}^np_xe_x(w_x)$. For the binary case (i.e. $n=2$) the success probability reads as $P^{\Omega}_S=p_2+\sup_{e\in\mathcal{E}}[p_1e(\omega_1)-p_2e(\omega_2)]$, which takes the form $ P^Q_S=\frac{1}{2}(1+||p_1\rho_1-p_2\rho_2||_1)$ in QT \cite{Helstrom'69}, where $||A||_1$ is the trace norm for the operator $A$ defined as $||A||_1:=\mbox{tr}|A|=\mbox{tr}\sqrt{A^{\dagger}A}$. This bound is generally known as the Helstrom bound and can also be rewritten as
\begin{equation}
P^Q_S=\frac{1}{2}\left(1+\sqrt{1-4p_1p_2\mbox{tr}(\rho_1\rho_2)}\right).\label{Helstrom-Q}
\end{equation}
The measurement obtaining the optimal success probability is called Helstrom measurement. 
\par
Recently, Schmid and Spekkens have derived an upper bound of the success probability of minimal error state discrimination in any experiment that allows a preparation non-contextual description \cite{Schmid'17}.  
The assumption of preparation non-contextuality assures unique probability distribution over the ontological space for any two operationally equivalent preparation procedures.
In any preparation non-contextual model underlying QT the success probability of minimal error discrimination of two quantum states $\rho_1,\rho_2$ given with a prior probability distribution $\{p_1,p_2\}$ is bounded by
\begin{equation}
P^{NC}_S\le 1-\min\{p_1,p_2\}\mbox{tr}(\rho_1\rho_2).\label{Helstrom-NC}
\end{equation}
Note that, quantum Helstrom bound (given in Eq.(\ref{Helstrom-Q})) is always greater than or equal to the non-contextual bound. For non-trivial case (i.e., when $\mbox{tr}[\rho_1\rho_2]\neq0$) it is strictly greater. Clearly, this shows an operational manifestation of preparation contextuality. Implication of preparation contextuality, in other kind of operational tasks, has also been demonstrated recently \cite{Spekkens'09,Banik'15,Ambainis'16}. We will show preparation contextual advantage in minimal error state discrimination can also be obtained in GPTs other than QT. Before that let us present a more interesting result: that in the bipartite scenario how one can ennoble this non-contextual bound of state discrimination to establish nonlocality of bipartite quantum states.

\emph{Result.} Consider the following guessing game between two spatially separated players, Alice and Bob. The players share several copies of a bipartite quantum state $\rho_{AB}$. Alice is asked to prepare Bob's system in two different ensembles $q_1\rho_1+(1-q_1)\sigma_1$ and $q_2\rho_2+(1-q_2)\sigma_2$ respectively, by performing suitable measurements on her part. NS constraint implies $q_1\rho_1+(1-q_1)\sigma_1=q_2\rho_2+(1-q_2)\sigma_2=\mbox{tr}_A(\rho_{AB})$. Bob performs a two outcome POVM $\{E_1,E_2\}$ to maximize the following payoff function
\begin{equation}
\mathcal{F}:=\frac{q_1}{q_1+q_2}\mbox{tr}(E_1\rho_1)+\frac{q_2}{q_1+q_2}\mbox{tr}(E_2\rho_2).\label{game}
\end{equation}  
Note that, the payoff function $\mathcal{F}$ is the success probability of minimal-error discrimination of the quantum states $\rho_1$ and $\rho_2$ given with a prior probability distribution $\{p_i=q_i/(q_1+q_2)\}_{i=1}^2$. It therefore cannot overthrow the corresponding Helstrom bound. The quantum payoff thus always satisfy the following inequality, $\mathcal{F}\le1/2(1+\sqrt{1-4q_1q_2\mbox{tr}(\rho_1\rho_2)/(q_1+q_2)^2}):=\mathcal{F}^{opt}$. The optimal value is achieved when Bob performs a measurement $\{E_1,E_2\}$ satisfying the condition, $\mbox{tr}(E_i\sigma_i)=0$, for $i=1,2$ \cite{Bae'11}.

Obtaining a higher payoff depends on two facts: (i) Bob's ability to perform the proper measurement (Helstrom measurement for the optimal case) and (ii) Alice's ability to prepare the two required ensembles at Bob's end by performing suitable measurement on her part of the bipartite state $\rho_{AB}$. These raise the question which bipartite states will best serve the purpose. It is evident that Alice cannot remotely prepare the required decompositions of Bob's marginal state by performing measurement on her part when they share a product state. However, if the shared state is correlated (non-product) then, using the correlation, Alice may  approximately prepare the desired decompositions. Still, inexact preparation limits the payoff value. At this point, it is intriguing to ask: What if they share unsteerable state?
Before arriving to the answer, let us digress on the concept of steering a bit. Steering was first introduced by Schr\"{o}dinger in the early days of QT \cite{Schrodinger'36}, and a more rigorous formulation of this particular concept is given recently by Wiseman et al \cite{Wiseman'07}. For an astute understanding of steering, consider that Alice can perform different measurements $M_a=\{E_{x|a}|E_{x|a}\ge 0, \sum_xE_{x|a}=\mathbf{1}\}$ on her part of a bipartite quantum state $\rho_{AB}$, here the index $a$ denotes Alice's choice of measurements. Upon obtaining the POVM effect $E_{x|a}$, she remotely prepares Bob's system in the conditional state $\sigma_{x|a}=\mbox{tr}_A[(E_{x|a}\otimes\mathbf{1})\rho_{AB}]/P(x|a)$, where $P(x|a)=\mbox{tr}[(E_{x|a}\otimes\mathbf{1})\rho_{AB}]$. The collection $\{P(x|a),\sigma_{x|a}\}$ is referred to as an assemblage which completely characterizes the scenario and in accordance with no-signaling it satisfies the condition $\sum_xP(x|a)\sigma_{x|a}=\mbox{tr}_A(\rho_{AB}),\forall ~a$. The bipartite state $\rho_{AB}$ is called unsteerable if there exists a fine grained ensemble of states $\{p(\lambda),\sigma_{\lambda}|~\sum_{\lambda}p(\lambda)\sigma_{\lambda}=\mbox{tr}_A(\rho_{AB})\}$ such that $\sigma_{x|a}=\sum_{\lambda}p(\lambda)p(x|a,\lambda)\sigma_{\lambda}, \forall~x,a$ \cite{Wiseman'07}, i.e., whenever Alice and Bob share an unsteerable state then all the different ensembles of Bob's marginal state that Alice is able to  prepare can actually be mapped from a fine grained ensemble of states. This fine grained ensemble is also called as local hidden state model. Thus for unsteerable states, different decompositions of Bob's state prepared by Alice, allow a preparation non-contextual description in terms of local hidden state model. Since preparation non-contextuality limits the success probability of binary state discrimination, hence the payoff function of the above guessing game is  bounded by the same non-contextual(NC) bound whenever Alice and Bob share some unsteerable states, i.e., 
\begin{equation}
\mathcal{F}^{NC}\le 1-\min\left\{\frac{q_1}{q_1+q_2},\frac{q_2}{q_1+q_2}\right\}\mbox{tr}(\rho_1\rho_2).\label{steer}
\end{equation} 
A payoff value larger than the above bound certifies steerability of the shared state. Indeed, a larger payoff establishes nonlocality of the bipartite state. This is because non-contextual bound in Eq.(\ref{Helstrom-NC}) has been derived under the assumption of preparation non-contextuality only. No particular kind of ontological variable is considered. Therefore, whenever the success probability (i.e. payoff value) of guessing surpasses the corresponding NC bound, it excludes the possibility of not only a hidden state description (fine grained ensemble) of Bob's  assemblage (remotely prepared by Alice) but also excludes the possibility of any kind of preparation non-contextual description. Hence, it establishes nonlocal behavior of the shared state. A similar argument of  nonlocality was presented first in Ref.\cite{Harrigan'10} under the assumption that the ontological model underlying QT is $\psi$-ontic in nature. It has further been extended to a larger family of ontological models called non maximally $\psi$-epistemic models \cite{Leifer'13,Banik'14}. Compared to these arguments, novelty of the present one is that it is not restricted to any particular class of ontological models.
Since the present method reveals nonlocality through an operational task it can be tested experimentally. On the other hand, in comparison to any Bell test, it involves less number of measurements, particularly, two by Alice and one by Bob.   
\par
For an explicit example, consider that Alice is supposed to prepare two ensembles $q|\psi\rangle\langle\psi|+(1-q)\sigma_{\psi}$ and $q|\phi\rangle\langle\phi|+(1-q)\sigma_{\phi}$ at Bob's end, where $|\psi\rangle=|0\rangle$ and $|\phi\rangle=a|0\rangle+b|1\rangle$, with $a,b\in\mathbb{R}$ and $a^2+b^2=1$. The NS condition implies $q|\psi\rangle\langle\psi|+(1-q)\sigma_{\psi}=q|\phi\rangle\langle\phi|+(1-q)\sigma_{\phi}=\rho_B=\mbox{tr}(\rho_{AB})$. The optimal value of the payoff function $\mathcal{F}$, in this case, turns out to be $\mathcal{F}^{opt}=1/2(1+b)$, while the corresponding NC bound is given by $\mathcal{F}^{NC}=1-a^2/2$. Bob's optimal measurement $\{E_1,E_2\}$ satisfies the condition $\mbox{tr}(E_1\sigma_{\psi})=0=\mbox{tr}(E_1\sigma_{\phi})$, which, in turn, implies $E_1=|\chi\rangle\langle\chi|$ and $\sigma_{\psi}=|\chi^{\perp}\rangle\langle\chi^{\perp}|$, with $|\chi\rangle=\cos\theta|0\rangle-\sin\theta|1\rangle$, $\tan 2\theta=b/a$. The choice of $E_2$ and $\sigma_{\phi}$ are also get fixed accordingly. Consequently, the optimal bipartite state $\rho_{AB}$ also gets fixed, i.e., $\rho_{AB}=|\alpha\rangle_{AB}\langle\alpha|$, with $|\alpha\rangle_{AB}=\sqrt{\beta_1}|\beta'_1\rangle_A\otimes|\beta_1\rangle_B+\sqrt{\beta_2}|\beta'_2\rangle_A\otimes|\beta_2\rangle_B$. Here, $\beta_1=(1+b-\sqrt{1+3b^2})/(2+2b)$, and $\beta_2=(1+b+\sqrt{1+3b^2})/(2+2b)$ are the eigenvalues of $\rho_B$ with respective eigenstates $|\tilde{\beta_1}\rangle:=((-1-b^2+\sqrt{1+3b^2})/ab,1)$ and $|\tilde{\beta_2}\rangle:=(-1-b^2-\sqrt{1+3b^2})/ab,1)$ (tilde represents that the states are un-normalized); $\{|\beta'_1\rangle,|\beta'_2\rangle\}$ is some orthonormal basis on Alice side. Gisin-Hughston-Jozsa-Wootters theorem  guarantees the existence of measurements on Alice's side to prepare the required assembles \cite{Gisin'89,Hughston'93}. Note that for a given pair of $|\psi\rangle$ and $|\phi\rangle$, one non-maximally pure entangled state $|\alpha\rangle_{AB}$ provides the optimal success, and varying the state $|\phi\rangle$ (i.e., varying $a$ and $b$) it is easy to show that all non-maximally entangled pure states provide optimal success probability in one of the above guessing games. In Bell scenario analogous feature is observed for tilted-CHSH game \cite{Acin'12}. The proposed guessing game thus constitutes an empirical entanglement verification scheme, which in comparison to all such schemes, like state tomography, constructing entanglement witness operator, or observing violation of some steering or nonlocality inequality, uses less number of entangled pairs.
\par
\begin{figure}[b!]
 \centering
  \includegraphics[height=6cm,width=7cm]{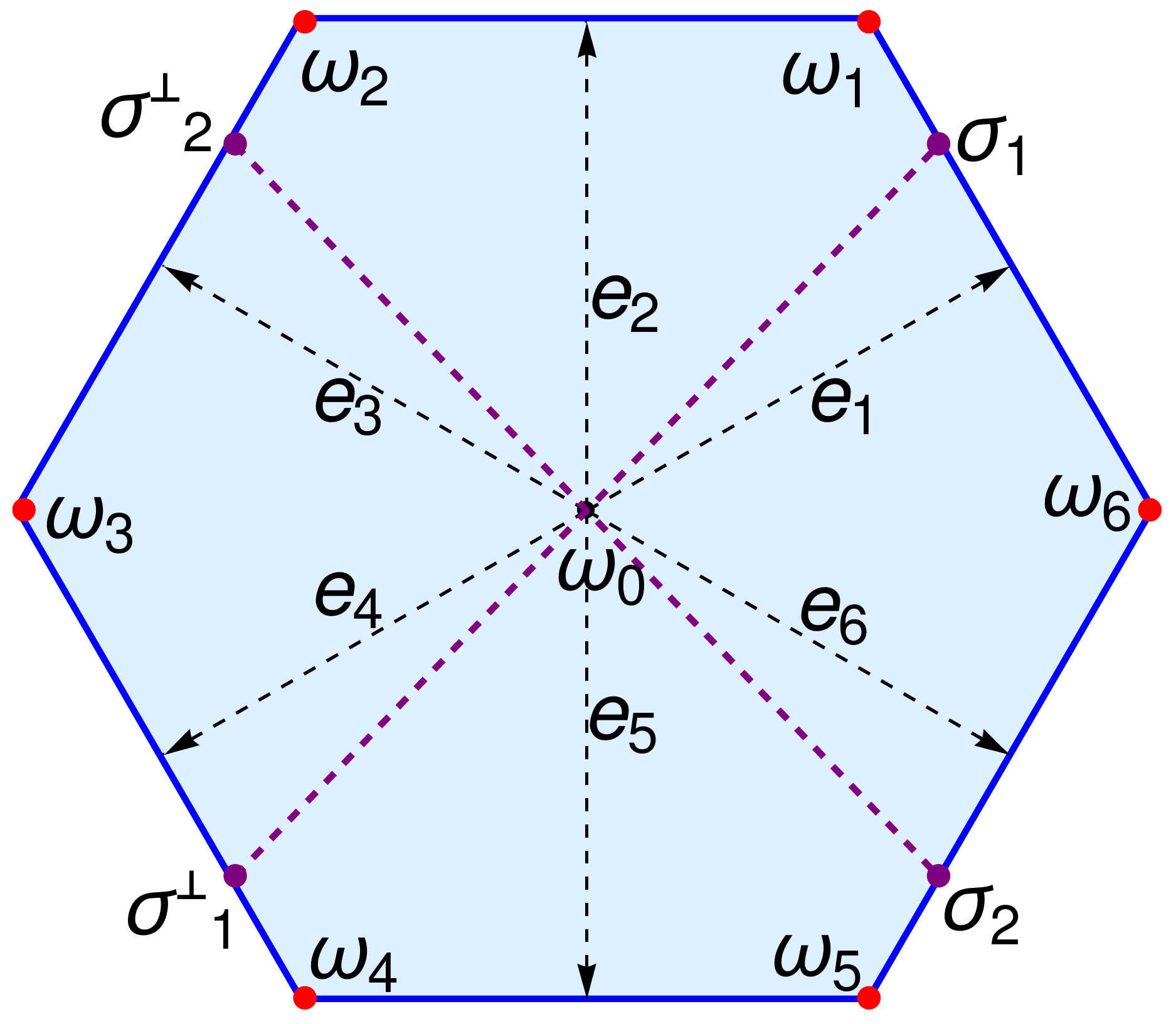}
 \caption{(Color online) Hexagon state space $\Omega_6$ ($z=1$ plane). Red dots denote six extremal states $\{\omega_i\}_{i=1}^6$ and black dashed lines with arrow-head denote six extremal effects $\{e_i\}_{i=1}^6$. Center point is the completely mixed state $\omega_0$. Note that the extremal $e_1$ gets filtered deterministically on all the states $q\omega_1+(1-q)\omega_6$, with $q\in[0,1]$. On the other hand $\omega_6$ deterministically filters both the extremal effects $e_1$ and $e_6$. This is in stark contrast with QT, where a pure state $|\psi\rangle$ deterministically filters only one rank one projector, i.e., $|\psi\rangle\langle\psi|$ and conversely $|\psi\rangle\langle\psi|$ gets filtered deterministically only on one pure states, i.e., $|\psi\rangle$. Here, we are interested in minimal state discrimination between the states $\sigma_1=p\omega_1+(1-p)\omega_6$ and $\sigma_2=p\omega_5+(1-p)\omega_6$.}\label{fig1}
\end{figure}
Interestingly, we find that preparation contextual advantage in minimal error state discrimination can also be observed in other GPTs. For example, consider the hexagon model (see Fig.\ref{fig1}). It consists of six pure states given by $\omega_j=(k\cos(2\pi j/6),k\sin(2\pi j/6),1)\in\mathbb{R}^3$, where $k=\sqrt{\sec(\pi/6)}$ and $j=1,...,6$ \cite{Janotta'11}. 
Consider two states: $\sigma_1=p\omega_1+(1-p)\omega_6$ 
and $\sigma_2=p\omega_5+(1-p)\omega_6$. 
Under the assumption of preparation non-contextuality the success probability of minimal error discrimination of these two states (given with prior probabilities $p_1,p_2$) is bounded by
$G(\sigma_1,\sigma_2)\le 1-\min\{p_1,p_2\}\left(1-\frac{p}{2}\right)$.
Whereas, the theory allows a success probability $P_S^{\Omega_6}=p_1e_2(\sigma_1)+p_2e_5(\sigma_2)=(1+p)/2$, which may supersede the corresponding non-contextual bound (see appendix).
Similar kind of advantage is possible in other polygon state spaces $\Omega_n$ (with $n=8,10,12,...$) \cite{Janotta'11}.

\emph{Discussion.} While comparing with Bell's theorem, one has to take a closer look at the method of revealing nonlocality proposed in this article. Given a joint probability distribution generated from some measurements at different spatial locations, Bell's inequality tests whether it is nonlocal or not irrespective of the rules and structure of the physical theory involved. This method of revealing nonlocality (if any) will not work for correlations generated from the three measurements scenario. But if there are further information regarding the structure and rules about state and measurement of the operational theory (physical or hypothetical), then there lies a possibility to demonstrate the nonlocality of the theory by exploiting those particular features of the theory. The method we describe here, is an example of such an approach. Our method of revealing nonlocality is stronger than similar kind of arguments proposed in Ref.\cite{Harrigan'10,Leifer'13}. Whereas those arguments work for special classes of ontological models, our method, like Bell's argument, works for every ontological models. Moreover, in terms of detection loophole as well as generating random measurement directions, our scheme is more robust than Bell type tests as it requires only one measurement on Bob's side. However, in contrast to Bell's approach, the asymmetry in our method, at least logically, does not negate the possibility for nonlocality to be one-way as well as non-monogamous, which demands further inquisitive research in this direction.             

State discrimination is a very primitive protocol in quantum information theory. Its relation to other fundamental results, such as no-cloning, no-signaling and its practical importance in a wide range of quantum information applications have been  extensively studied. Here we have pointed out application of this protocol in another very important task: certifying non-classicality of shared bipartite state, in particular, empirical entanglement verification protocol. A suitable multipartite generalization of our method may be useful to certify presence of genuine entanglement. We have discussed examples of generalized probability theories other than quantum theory which also violates non-contextual upper bound. Our guessing game can be presented in the framework of generalized probabilistic theories, which in turn gives opportunity to verify presence of non-classical correlations in those theories. Whether this method can be used to compare strength of correlations in different theories may be an interesting question of further research.  

{\bf Acknowledgments.} SSB, AR, AM, and GK thank Tamal Guha for useful discussion. SK would like to acknowledge visit at The Institute of Mathematical Sciences, Chennai, where this work has been done. AM acknowledges support from the CSIR project 09/093(0148)/2012-EMR-I.

\begin{widetext}
	
\section*{Appendix}
\emph{Contextual advantage of state discrimination in GPT.} As already discussed, in the GPT framework state space $\Omega$ forms a convex set embedded in some real vector space. The extreme point of $\Omega$ are called pure states, while the rests are called mixed states that allow decompositions in terms of pure states. If the state space, unlike in the classical case, is not a simplex, then a mixed states allow more than one decompositions in terms of pure states. Operationally, these different decompositions represent different preparation procedures of the same mixed state. All these different preparations are operationally equivalent in the sense that no measurement can distinguish these preparations. Let us denote this operational equivalence by $`\approx'$. Therefore $\omega_{I}=\sum_ip_i\omega_i$ and $\omega_{II}=\sum_jq_j\omega'_j$ are operationally equivalent, i.e., $\omega_{I}\approx\omega_{II}$ \emph{iff} $e(\omega_{I})=e(\omega_{II})$, for all the effects $e\in\mathcal{E}$. An ontological model underlying this GPT will be called preparation non-contextual if operationally equivalent preparation gives equivalent probability distribution on the ontic state space, i.e., $\mu_{\omega_{I}}(\lambda)=\mu_{\omega_{II}}(\lambda)$, whenever $\omega_{I}\approx\omega_{II}$. 
\begin{table}[b!]
	\begin{center}
		\begin{tabular}{|m{2.5em}||m{2.5em}|m{4em}| m{4em}| m{4em}|} 
			\hline
			& $\sigma_1$ & $\sigma_2$ & $\sigma_1^{\perp}$ & $\sigma_2^{\perp}$ \\  [5pt]
			\hline\hline
			$~~e_1$ & $~~~1$ & $~~~~\mathcal{C}_{\Omega_6}$ & $~~~~~0$ & $~~1-\mathcal{C}_{\Omega_6}$ \\  [5pt]
			\hline
			$~~e_6$ & $~~~\mathcal{C}_{\Omega_6}$ & $~~~~~1$ & $~~1-\mathcal{C}_{\Omega_6}$ & $~~~~~0$ \\   [5pt] 
			\hline
			$~~e_2$ & $~~~\mathcal{S}_{\Omega_6}$ & $~~1-\mathcal{S}_{\Omega_6}$ & $~~1-\mathcal{S}_{\Omega_6}$ & $~~~~\mathcal{S}_{\Omega_6}$ \\   [5pt] 
			\hline
		\end{tabular}
		\caption{Outcome probabilities of the effect $e_1,e_6,e_2$ on the states $\{\sigma_i, \sigma_i^{\perp}\}_{i=1}^2$. Here $\mathcal{C}_{\Omega_6}=1-p/2$ and $\mathcal{S}_{\Omega_6}=(1+p)/2$.}
	\end{center}  
\end{table}
\par
Consider the hexagon model (see Fig.\ref{fig1} in the article). Six pure states given by $\omega_j=(k\cos(2\pi j/6),k\sin(2\pi j/6),1)\in\mathbb{R}^3$, where $k=\sqrt{\sec(\pi/6)}$ and $j=1,...,6$ \cite{Janotta'11-a}. Six pure effects are $e_j:=1/2(k\cos(2j-1)\pi/6,k\sin(2j-1)\pi/6,1)\in\mathbb{R}^3$, and outcome probability rule is specified by standard $\mathbb{R}^3$ inner product, i.e., $e(\omega)=e.\omega$.
Consider the states: $\sigma_1=p\omega_1+(1-p)\omega_6$$\sigma_1^{\perp}=p\omega_4+(1-p)\omega_3$, 
and $\sigma_2=p\omega_5+(1-p)\omega_6$, and  $\sigma_2^{\perp}=p\omega_2+(1-p)\omega_3$, and three measurements: $M_1=\{e_1,e_4\}$, $M_2=\{e_3,e_6\}$, and $M_3=\{e_2,e_5\}$. The outcome probabilities of the considered states on the considered measurement is listed in Table-I.
\par
Any ontological model $(\Lambda,\mu,\xi)$ underlying $\Omega_6$ must reproduce the operational predictions listed in the table. Therefore we have, 
\begin{subequations}\label{repro}
\begin{align}
\int_{\lambda\in\Lambda}d\lambda\mu_{\sigma_1}(\lambda)\xi_{e_1|M_1}(\lambda)=1,\label{repro-a}\\ \int_{\lambda\in\Lambda}d\lambda\mu_{\sigma_1^{\perp}}(\lambda)\xi_{e_1|M_1}(\lambda)=0,\label{repro-b}\\ \int_{\lambda\in\Lambda}d\lambda\mu_{\sigma_1}(\lambda)\xi_{e_6|M_2}(\lambda)=\mathcal{C}_{\Omega_6},\label{repro-c}\\ \int_{\lambda\in\Lambda}d\lambda\mu_{\sigma_1}(\lambda)\xi_{e_2|M_3}(\lambda)=\mathcal{S}_{\Omega_6}.\label{repro-d}
\end{align}
\end{subequations}
Let us denote $\Lambda_{\sigma}:=\{\lambda\in\Lambda|\mu_{\sigma}(\lambda)>0\}$. Since $\int_{\lambda\in\Lambda_{\sigma}}d\lambda\mu_{\sigma_1}(\lambda)=1$, hence from Eq.(\ref{repro-a}) we can say $\xi_{e_1}(\lambda)=1,~\forall~\lambda\in\Lambda_{\sigma_1}$. To satisfy Eq.(\ref{repro-b}) we have $\Lambda_{\sigma_1}\cap\Lambda_{\sigma_1^{\perp}}=\emptyset$. With similar reasoning $\Lambda_{\sigma_2}\cap\Lambda_{\sigma_2^{\perp}}=\emptyset$. Please note that this is unlike quantum mechanics: in quantum mechanics such relation must hold when two states are orthogonal with Hilbert-Schmidt inner product \cite{self,Pusey'12}, but here $\sigma_i$ and $\sigma_i^{\perp}$ are not orthogonal with respect to standard $\mathbb{R}^3$ inner product.
\par
In the ontological model the response functions are not assumed to be deterministic in general. But if we assume that the ontological model underlying $\Omega_6$ is preparation contextual then we have $\xi_{e_i}(\lambda)\in\{0,1\}$, for $i=1,...,6$. To see this, consider a mixed state $\omega_0$ which is equal mixture of $\sigma_1$ and $\sigma_1^{\perp}$, i.e., $\omega_0=\frac{1}{2}\sigma_1+\frac{1}{2}\sigma_1^{\perp}$. This we can say to be the completely mixed state (in Fig.\ref{fig1} the center point of the hexagon). Every state $\omega\in\Omega_6$ appears in some decomposition of $\omega_0$. By the assumption of preparation non-contextuality every such decomposition has the same distribution over ontic states. Thus, every ontic state in the support of the corresponding $\mu_{\omega}(\lambda)$ also appears in the support of $\mu_{\omega_0}(\lambda)$, so the full state space $\Lambda$ is equivalent to $\Lambda_{\omega_0}=\Lambda_{\sigma_1}\cup\Lambda_{\sigma_1^{\perp}}$. As already argued $\xi_{e_1}(\lambda)=1,~\forall~\lambda\in\Lambda_{\sigma_1}$ and $\xi_{e_1}(\lambda)=0,~\forall~\lambda\in\Lambda_{\sigma_1^{\perp}}$, which proves the claim that $e_1$ has deterministic response over the ontic states. Similar argument holds for other $e_i$'s.
\begin{figure}[t!]
	\centering
	\includegraphics[height=5cm,width=8cm]{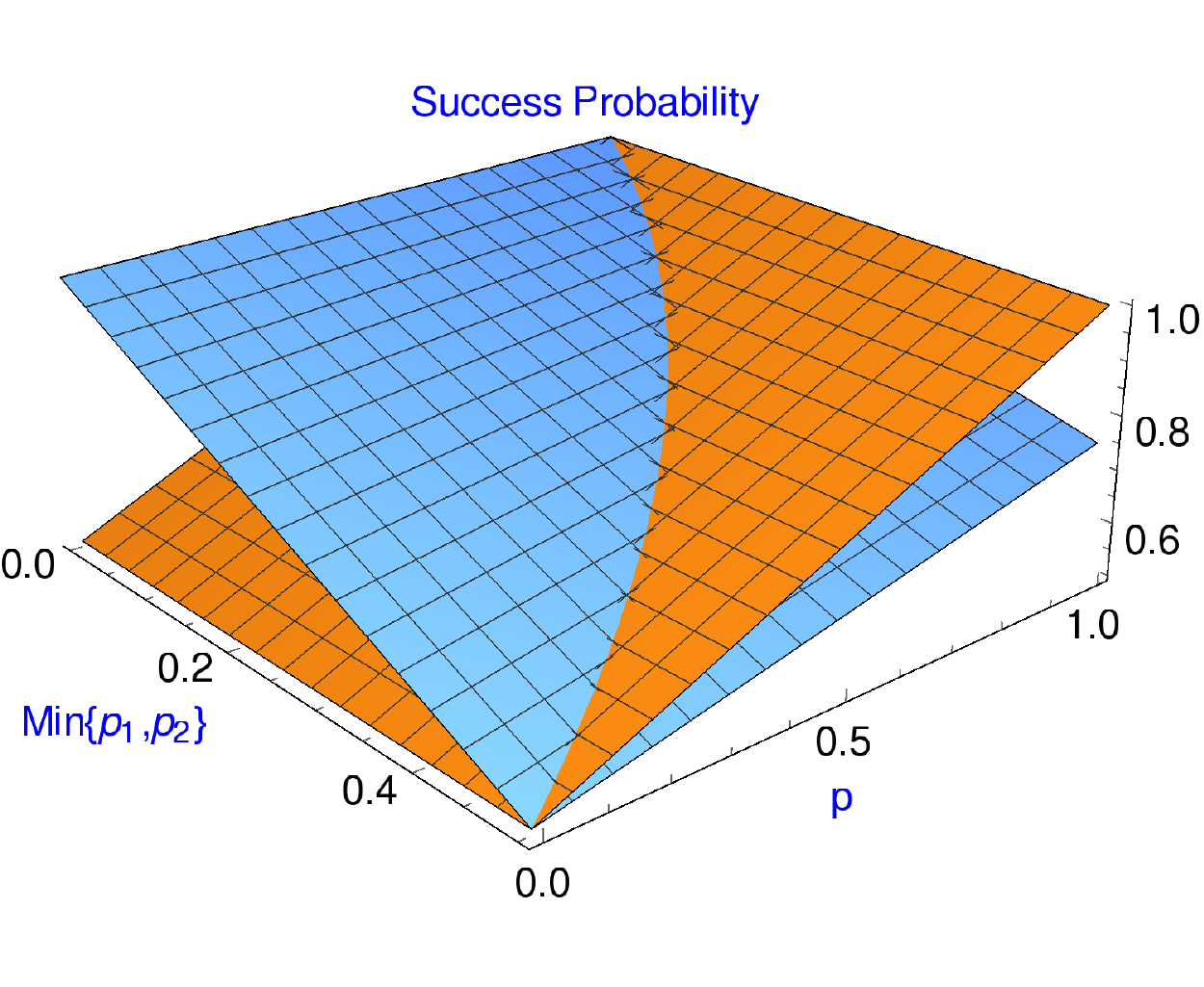}
	\caption{(Color on-line) Preparation contextual advantage of minimal error state discrimination in GPT with state space $\Omega_6$. Blue curve is the preparation non-contextual bound on the success probability $G(\sigma_1,\sigma_2)$ of the discriminating the states $\sigma_1$ and $\sigma_2$, given with a prior probability distribution $\{p_1,p_2\}$. Orange curve denotes the success probability $P^{\Omega_6}_S$ obtained while performing measurement $M_3=\{e_2,e_5\}$.}\label{fig2}
\end{figure}
\par
Suppose, a classical variable $\lambda$ is sampled from one of two overlapping probability distributions, $p(\lambda|a)$ and $p(\lambda|b)$. On average, the success probability probability of guessing which of the two distributions $\lambda$ is drawn from is given by (see Ref.\cite{Schmid'17-a} for more elucidation),  
\begin{eqnarray}
G(a,b)=1-\int_{\lambda}d\lambda\min\{p(\lambda|a)p(a),p(\lambda|b)p(b)\},\nonumber\\
\le 1-\min\{p(a),p(b)\}\int_{\lambda}d\lambda\min\{p(\lambda|a),p(\lambda|b)\}.
\end{eqnarray}
According to this formula, if two states $\sigma_1,\sigma_2$ are given with  prior probabilities $p_1,p_2$, then successful discrimination probability of this two state is bounded by,
\begin{eqnarray}
G(\sigma_1,\sigma_2)\le 1-\min\{p_1,p_2\}\int_{\lambda}d\lambda\min\{\mu_{\sigma_1}(\lambda),\mu_{\sigma_2}(\lambda)\}.\label{PNC-GPT}
\end{eqnarray}
Note that $\mathcal{C}_{\Omega_6}=\int_\lambda d\lambda\mu_{\sigma_1}(\lambda)\xi_{e_6}(\lambda)$. Since, in preparation non-contextual model, $\xi_{e_6}(\lambda)=1,~\forall~\lambda\in\Lambda_{\sigma_2}$ and $\xi_{e_6}(\lambda)=0,~\forall~\lambda\in\Lambda_{\sigma_2^{\perp}}$, we therefore can write $\mathcal{C}_{\Omega_6}=\int_{\lambda\in\Lambda_{\sigma_2}} d\lambda\mu_{\sigma_1}(\lambda)$. Also note that $\frac{1}{2}\mu_{\sigma_1}(\lambda)+\frac{1}{2}\mu_{\sigma_1^{\perp}}(\lambda)=\frac{1}{2}\mu_{\sigma_2}(\lambda)+\frac{1}{2}\mu_{\sigma_2^{\perp}}(\lambda)=\mu_{\omega_0}(\lambda)$ and $\Lambda_{\sigma_i}\cap\Lambda_{\sigma_i^{\perp}}=\emptyset$. This implies that, $\forall~\lambda\in\Lambda_{\sigma_1}\cap\Lambda_{\sigma_2}$, $\mu_{\sigma_1}(\lambda)=\mu_{\sigma_2}(\lambda)=2\mu_{\omega_0}(\lambda)$. Hence we have $\min\{\mu_{\sigma_1}(\lambda),\mu_{\sigma_2}(\lambda)\}=\mu_{\sigma_1}(\lambda)=\mu_{\sigma_2}(\lambda)$ for all $\lambda\in\Lambda_{\sigma_1}\cap\Lambda_{\sigma_2}$ and zero everywhere else. Consequently, $\mathcal{C}_{\Omega_6}=\int_{\lambda\in\Lambda_{\sigma_2}} d\lambda\mu_{\sigma_1}(\lambda)=\int_{\lambda\in\Lambda_{\sigma_2}} d\lambda\min\{\mu_{\sigma_1}(\lambda),\mu_{\sigma_2}(\lambda)\}=\int_{\lambda} d\lambda\min\{\mu_{\sigma_1}(\lambda),\mu_{\sigma_2}(\lambda)\}$. Accordingly Eq.(\ref{PNC-GPT}) becomes,
\begin{eqnarray}
G(\sigma_1,\sigma_2)\le 1-\min\{p_1,p_2\}\left(1-\frac{p}{2}\right), \label{eq8}
\end{eqnarray}
whereas the theory allows a success probability $P_S^{\Omega_6}=p_1e_2(\sigma_1)+p_2e_5(\sigma_2)=(1+p)/2$,
whenever the measurement $M_3=\{e_2,e_5\}$ is performed to discriminate the given pair of states. 
Clearly, $P_S^{\Omega_6}$ can supersede the non-contextual bound of Eq.(\ref{eq8}) (see Fig.\ref{fig2}). This establishes preparation contextual advantage in state discrimination in a GPT other than quantum theory. 
Similar kind of advantage is possible in other polygon state spaces $\Omega_n$ (with $n=8,10,12,...$), where extremal states are $\omega_i:=(k_n\cos(2\pi j/n),k_n\sin(2\pi j/n),1)\in\mathbb{R}^3$, with $j=1,2,...n$ and $k_n=\sqrt{\sec(\pi/n)}$; and the extremal effects are $\omega_i:=1/2(k_n\cos(2j-1)\pi/n,k_n\sin(2j-1)\pi/n,1)$ \cite{Janotta'11-a}.
\end{widetext}


\begin{thebibliography}{99}
\bibitem{self} Depending on the fact whether non orthogonal pure states give overlapping /non-overlapping distribution over ontological space, the ontological models are classified into types-- $\psi$-epistemic/ $\psi$-ontic. Recently under a assumption called preparation independence it has been shown that $\psi$-epistemic model can not reproduce the quantum statistics \cite{Pusey'12}.
\end{thebibliography}
\end{document}